\documentstyle[erice97,psfig,epsfig,11pt]{article}
\def\bnue{\hbox{$\bar\nu_e$ }}  
\def\bnum{\hbox{$\bar\nu_\mu$ }}

\newcommand {\ignore}[1]{}

\newcommand{\bc}{\begin{center}}
\newcommand{\ec}{\end{center}}

\def\ifmath#1{\relax\ifmmode #1\else $#1$\fi}
\def\half{\ifmath{{\textstyle{1 \over 2}}}}

\def\3quarter{{\textstyle{3 \over 4}}}

\def\ra{\rightarrow}

\def\lf{\leaders\hbox to 1em{\hss.\hss}\hfill}
\def\e6{$E(6)$}
\def\10{$SO(10)$}
\def\21{$SU(2) \otimes U(1) $}
\def\lr{$SU(2)_L \otimes SU(2)_R \otimes U(1)$}
\def\422{$SU(4) \otimes SU(2) \otimes SU(2)$}
\def\321{$SU(3) \otimes SU(2) \otimes U(1)$}
\def\ne{\hbox{$\nu_e$ }}
\def\nm{\hbox{$\nu_\mu$ }}
\def\nt{\hbox{$\nu_\tau$ }}

\def\mnt{\hbox{$m_{\nu_\tau}$ }}

\def\neus{\hbox{neutrinos }}

\def\neu{\hbox{neutrino }}

\def\eq#1{{eq. (\ref{#1})}}

\def\fig#1{{Fig. (\ref{#1})}}

\def\VEV#1{\left\langle #1\right\rangle}
\let\vev\VEV

\def\lsim{\raise0.3ex\hbox{$\;<$\kern-0.75em\raise-1.1ex\hbox{$\sim\;$}}}
\def\gsim{\raise0.3ex\hbox{$\;>$\kern-0.75em\raise-1.1ex\hbox{$\sim\;$}}}
\def\half{{1\over 2}}
\def\beq{\begin{equation}}
\def\eeq{\end{equation}}
\def\bef{\begin{figure}}
\def\eef{\end{figure}}
\def\bet{\begin{table}}
\def\eet{\end{table}}
\def\bea{\begin{eqnarray}}
\def\ba{\begin{array}}
\def\ea{\end{array}}
\def\bi{\begin{itemize}}
\def\ei{\end{itemize}}
\def\ben{\begin{enumerate}}
\def\een{\end{enumerate}}
\def\ra{\rightarrow}

\def\eea{\end{eqnarray}}
\def\apj#1#2#3{          {\it Astrophys. J. }{\bf #1} (19#2) #3}

\def\ib#1#2#3{           {\it ibid. }{\bf #1} (19#2) #3}
\def\nat#1#2#3{          {\it Nature }{\bf #1} (19#2) #3}
\def\nps#1#2#3{        {\it Nucl. Phys. B (Proc. Suppl.) }{\bf #1} (19#2) #3} 
\def\np#1#2#3{           {\it Nucl. Phys. }{\bf #1} (19#2) #3}
\def\pl#1#2#3{           {\it Phys. Lett. }{\bf #1} (19#2) #3}
\def\pr#1#2#3{           {\it Phys. Rev. }{\bf #1} (19#2) #3}

\def\prl#1#2#3{          {\it Phys. Rev. Lett. }{\bf #1} (19#2) #3}

\def\n.c.#1#2#3{         {\it Nuovo Cim. }{\bf #1} (19#2) #3}
\def\r.n.c.#1#2#3{       {\it Riv. del Nuovo Cim. }{\bf #1} (19#2) #3}

\def\jetp#1#2#3{         {\it JETP }{\bf #1} (19#2) #3}
\def\mpl#1#2#3{          {\it Mod. Phys. Lett. }{\bf #1} (19#2) #3}

\def\ppnp#1#2#3{           {\it Prog. Part. Nucl. Phys. }{\bf #1} (19#2) #3}

\def\tp{these proceedings}

\def\ip{in preparation}

\begin{document}
\title{\Large  Recent Results on Neutrino Masses}
\author{Jos\'e W. F. Valle}
\address{Instituto de F\'{\i}sica Corpuscular 
- C.S.I.C.\\Departament de F\'{\i}sica Te\`orica, Universitat de
Val\`encia\\46100 Burjassot, Val\`encia,
Spain\\http://neutrinos.uv.es}
\maketitle
\abstracts{
I review the main options one has of introducing mass to neutrinos,
including broken R--parity models, as well as the constraints on
neutrino properties that follow from astrophysics, cosmology as well
as laboratory observations.  }

\section{Introduction}

One of the most unpleasant features of the Standard Model (SM) is that
the masslessness of neutrinos is not dictated by an underlying
principle, such as gauge invariance, in the case of the photon: the SM
simply postulates that neutrinos are massless and, as a result, most
of their properties are trivial.  Massless neutrinos would be
exceptional particles, since no other such fermions exist. If massive,
neutrinos would present another puzzle, of why are their masses so
much smaller than those of the charged fermions. The fact that
neutrinos are the only electrically neutral elementary fermions means
that they could be Majorana-type, the most fundamental one. This may
hold the key to why neutrino masses are small: they break lepton
number symmetry. This feature is encountered in many extensions of the
SM. However, at the present stage of affairs, theory alone cannot make
definite predictions for the magnitude of particle masses, and \neus
are no exception.  We simply do not know the scale responsible for
neutrino mass, nor the underlying mechanism. Last but not least, one
lacks a theory for the Yukawa couplings. In what follows I will
discuss the main options one has of introducing mass to neutrinos, and
over-view the present information that follows from astrophysics,
cosmology as well as laboratory observations.

\section{Models of Neutrino Mass}

One of the most popular approaches to generate neutrino masses is
Unification. Indeed, in trying to understand the origin of parity
violation in the weak interaction by ascribing it to a spontaneous
breaking phenomenon, in the same way as the W and Z acquire their
masses in the SM, one arrives at the so-called left-right symmetric
extensions \lr \cite{LR}, \422 \cite{PS} or \10 \cite{GRS}, in some of
which the masses of the light neutrinos are obtained by diagonalizing
the following mass matrix in the basis $\nu,\nu^c$
\begin{equation}
\left[\matrix{
 0 & D \cr
 D^T & M_R }\right] 
\label{SS} 
\end{equation} 
where $D$ is the standard \21 breaking Dirac mass term and $M_R =
M_R^T$ is the isosinglet Majorana mass associated to the violation of
the extended gauge symmetry. In the seesaw approximation, one finds
\beq M_{eff} = - D M_R^{-1} D^T \:.  
\label{SEESAW} 
\eeq 
In general, however, one also has a $\nu\nu$ term whose size is
expected to be also suppressed by the right-handed breaking
scale. This way one can naturally explain the smallness of
\neu masses.  Even though $M_R$ should be large, its magnitude heavily
depends on the model. As a result one can not make any solid
prediction for the light neutrino mass spectrum generated through the
seesaw mechanism, nor for the resulting pattern of mixing
\cite{2227}. In fact this freedom has been exploited in model building
in order to account for an almost degenerate \neu mass spectrum
presently suggested by solar and atmospheric \neu data as well as the
recent observations on structure formation \cite{DEG}.

Although very attractive, Unification is by no means the only way to
generate neutrino masses. There is a large diversity of other possible
schemes which do not require any new large mass scale. For example, it
is possible to start from an extension of the lepton sector of the \21
theory by adding a set of $two$ 2-component isosinglet neutral
fermions, denoted ${\nu^c}_i$ and $S_i$, to each generation. In this
case there is an exactly conserved L symmetry that keeps neutrinos
strictly massless, as in the SM, leading to the following form for the
neutral mass matrix (in the basis $\nu, \nu^c, S$)
\begin{equation}
\left[\matrix{
  0 & D & 0 \cr
  D^T & 0 & M \cr
  0 & M^T & 0 }\right] 
\label{MAT} 
\end{equation} 
This form has also been suggested in various theoretical models
\cite{WYLER}, including many of the superstring inspired models
\cite{SST}. In the latter case the zeros of \eq{MAT} arise due 
to the absence of Higgs fields to provide the usual Majorana mass
terms. Although neutrinos are {\sl massless} in this model, they are
{\sl mixed} in the leptonic charged current in a non-trivial way which
cannot be eliminated by field redefinitions.  Conventional neutrino
oscillation searches {\sl in vacuo} are insensitive to this
mixing. However, such neutrinos may resonantly convert in the dense
medium of a supernova \cite{massless0,massless}.  One can also
introduce non-zero masses in this model through a $\mu S S$ term that
could be proportional to the vacuum expectation value of a singlet
field $\sigma$ \cite{CON}.  In contrast to the seesaw scheme,
\neu masses would be directly proportional to $\VEV{\sigma}$,
$m_\nu \propto \vev{\sigma}$.  The massless \neu model characterized by
\eq{MAT} provides a conceptually simple and phenomenologically rich
extension of the Standard Model, which brings in the possibility that
a wide range of new phenomena be sizeable. These have to do with
neutrino mixing, universality, flavour and CP violation in the lepton
sector
\cite{BER,CP}, as well as direct effects associated with neutral heavy
lepton production at high energy colliders \cite{CERN}.  The rates for
loop-induced lepton flavour and CP non-conservation effects in this
model are precisely calculable \cite{BER,CP,3E} and can be large as
they are not restricted by \neu masses, only by weak universality
violation constraints, which are weaker. The only other example I know
of getting sizeable lepton flavour violation (LFV) without \neu masses
is the supersymmetric mechanism first described in ref. \cite{Hall}.
It is interesting to realize that LFV does not require \neu masses,
unlike total-L-violating processes, such as neutrino-less double beta
decay, see below.

\begin{figure}[t]
\centerline{\protect\hbox{\psfig{file=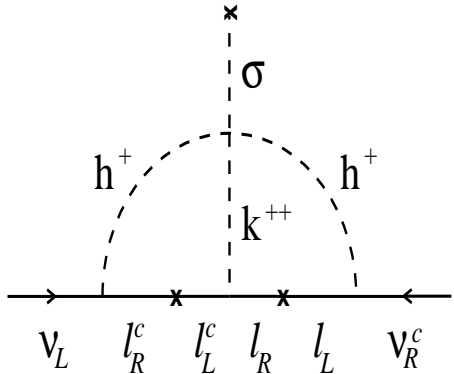,height=5cm,width=6cm}}}
\caption{Two-loop-induced Neutrino Mass. }
\label{2loop}
\end{figure}

There is also a large variety of {\sl radiative} schemes to generate
\neu masses. The prototype models of this type are the Zee model and
the model suggested by Babu \cite{zee.Babu88}. In these models lepton
number is explicitly broken, but it is easy to enforce the spontaneous
breaking of lepton number by adding a single scalar boson $\sigma$, as
in the version of ref. \cite{ewbaryo}. Note that the neutrino mass
arising from the diagram shown in \fig{2loop}
goes to zero as $\vev{\sigma} \to 0$.

\begin{figure}[t]
\centerline{\protect\hbox{\psfig{file=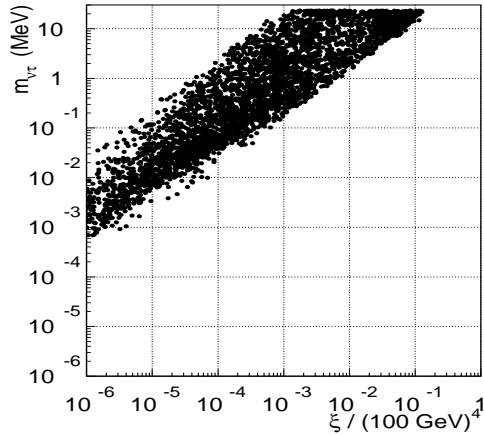,height=6cm,width=7cm}}}
\caption{Tau neutrino mass versus $\epsilon_3$ }
\label{mnt_xi_ev}
\end{figure}

\begin{figure}[t]
\centerline{\protect\hbox{\psfig{file=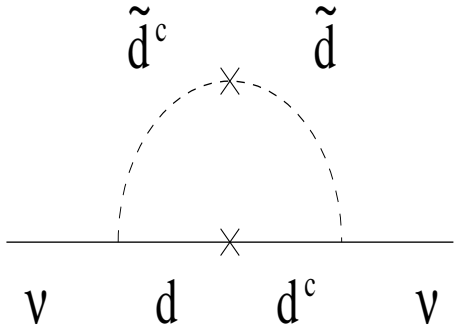,height=5cm,width=6cm}}}
\caption{Scalar loop contributions to neutrino masses  }
\label{mnrad}
\end{figure}

The seesaw and the radiative mechanisms of neutrino mass generation
may be combined. Supersymmetry with broken R-parity provides a very
elegant mechanism to do this, and can explain the origin of neutrino
mass, as well as mixings \cite{epsrad}. Here I focus on the simplest
unified supergravity version of the model with bilinear breaking of
R--parity, characterized universal boundary conditions for the soft
breaking parameters \cite{epsrad}. In this model the tau neutrino
(which is the fifth in the following matrix) acquires a mass, due to
the mixing between the \neu and the neutralinos given in the matrix
\cite{RPothers}
\begin{equation}
\left[\matrix{
M_1 & 0  & -\half g'v_1 & \half g'v_2 & -\half g'v_3 \cr
0   & M_2 & \half g v_1 & -\half g v_2 & \half g v_3 \cr
-\half g'v_1 & \half g v_1 & 0 & -\mu & 0 \cr
\half g'v_2 & -\half g v_2 & -\mu & 0 & \epsilon_3 \cr
-\half g'v_3 & \half g v_3 & 0 & \epsilon_3 & 0 
}\right]
\label{eq:NeutMassMat}
\end{equation}
This matrix is a truncation of the more complete one that one finds in
dynamical models with spontaneous breaking of R--parity and which
necessarily contain singlet neutrino superfields \cite{Romao92}.  Here
I focus on the simplest truncated model with explicit violation of
R--parity. Contrary to a popular misconception, the bilinear violation
of R--parity implied by the parameter $\epsilon_3$ is physical, and
can not be rotated away. One can show that by going to a basis in
which $\epsilon_3$ is eliminated from the superpotential one gets
exactly the same results as in the original un-rotated basis
\cite{DJV}.  A minimal supergravity version of this model contains 
only one extra free parameter in addition to those of the minimal
R--parity conserving supergravity model, as the $\epsilon_3$ and the
$v_3$ are related by a minimization condition \cite{epsrad}.  The
assumed universality of the soft breaking parameters implies that the
value of $\epsilon_3$ is induced radiatively, due to the effect of the
non-zero bottom quark Yukawa coupling $h_b$ in running the
renormalization group equations from the unification scale down to the
weak scale \cite{epsrad}.  This makes $\epsilon_3$ {\sl
calculable}. Thus \eq{eq:NeutMassMat} is analogous to a see-saw type
matrix \eq{SS}, in which the sector analogous to $M_R$ (neutralinos)
lies at the weak scale, while the r\^ole of the Dirac entry $D$ is
played by the $\epsilon_3$, which is, in a sense, a radiatively
induced quantity. From this point of view, the mechanism is a {\sl
hybrid} see-saw-like scheme, with the Majorana $\nu_{\tau}$ mass
induced by the mixing of neutrinos with {\sl Higgsinos} or {\sl
gauginos}. The \nt mass induced this way depends quadratically on an
effective parameter $\xi$ defined as $\xi \equiv (\epsilon_3 v_1 + \mu
v_3)^2$ and characterizing the violation of R--parity either through
gaugino ($v_3$) or Higgsino mixing ($\epsilon_3$).  In \fig{mnt_xi_ev}
we display the allowed values of $m_{\nu_{\tau}}$, which can be quite
low even for sizeable values of $\epsilon_3$, due to the possible
cancellation between the two terms in $\xi$. In the unified
supergravity version of the model with universal soft masses this
cancellation happens automatically. The value of $\xi$ becomes
calculable in terms of $h_b$, leading to a naturally suppressed \mnt
in the KeV range.
Notice that \ne and \nm remain massless in this approximation, even if
the other two parameters $\epsilon_1$ and $\epsilon_2$ are added to
complete the model for the first two families. This is an old result
related to the projective nature of the neutrino mass matrix in these
models \cite{arca}. The \nm will get mass either from scalar loop
contributions in \fig{mnrad} \cite{RPnuloops},
or by mixing with singlets in models with spontaneous breaking of
R-parity \cite{Romao92}. This leads to a solution to the solar \neu
puzzle via \ne to \nm conversions. In short, broken R--parity
supersymmetry provides an elegant mechanism for generating \neu
masses and mixings. It is important to notice that even when the
\neu masses are small, the corresponding R-parity violating 
effects can be large. For example the lightest neutralino typically
happens inside the detector, giving rise to remarkable signatures
at accelerator experiments such as LEP/LHC.

Other than the seesaw scheme, none of the above models requires a
large mass scale. In all of them one can implement the spontaneous
violation of the global lepton number symmetry leading to \neu masses
that scale {\sl directly} proportional to the lepton-number scale
$\vev{\sigma}$ or some positive power of it. Such low-scale Majoron
models are very attractive and lead to a rich phenomenology, as the
extra particles required have masses at scales that could be
accessible to present experiments. One remarkable example is the
possibility invisibly decaying Higgs bosons \cite{JoshipuraValle92}.

The above discussion illustrates the enormous wealth of
phenomenological possibilities in the neutrino sector. These reach
well beyond the realm of conventional neutrino experiments, including
also signatures that can be probed, though indirectly, at high energy
accelerators.  Unfortunately, as already mentioned, \neu masses are
not predicted by theory and it is up to observation to search for any
possible clue.  Given the theoretical uncertainties in predicting \neu
masses from first principles, one must turn to observation. Here the
information comes from laboratory, cosmology and astrophysics.

\section{Laboratory Limits }
\vskip .2cm

The only model-independent laboratory limits on \neu mass follow
purely from kinematics \cite{PDG96}:
\beq
\label{1}
m_{\nu_e} 	\lsim 5 \: \mbox{eV}, \:\:\:\:\:
m_{\nu_\mu}	\lsim 170 \: \mbox{keV}, \:\:\:\:\:
m_{\nu_\tau}	\lsim 18 \: \mbox{MeV}
\eeq 
The limit on the \ne mass comes from tritium beta decay
\cite{Lobashev} while further results from the Mainz experiment are
eagerly awaited. The \nm mass limit comes from PSI (90 \% C.L.)
\cite{psi}, with further improvement limited by the uncertainty in the
$\pi^-$ mass.  On the other hand, the best \nt mass limit now comes
from high energy LEP experiments
\cite{eps95} and may be substantially improved at a future tau-charm
factory \cite{jj}. 

The most stringent neutrino oscillation bounds come from reactor
experiments \cite{reactor} (\bnue - $\nu_x$ oscillations); from meson
factory oscillation experiments (KARMEN \cite{karmen}, LSND
\cite{lsnd}) and from high-energy accelerator experiments E531 and
E776 \cite{E531.E776} (\nm - \nt).  A recent search for \nm to \ne
oscillations by the LSND collaboration using \nm from $\pi^+$ decay in
flight \cite{lsndflight} is consistent with their previously reported
\bnum to \bnue oscillation evidence. 
However, a recent result from NOMAD rules out part of the LSND
region. The future lies in searches for oscillations using reactor or
accelerator \neu beams directed to far-out underground detectors, with good
prospects for Chooz and Palo Verde, as well as the long-baseline
experiments proposed at KEK, CERN and Fermilab.

\begin{figure}[t]
\centerline{\protect\hbox{\psfig{file=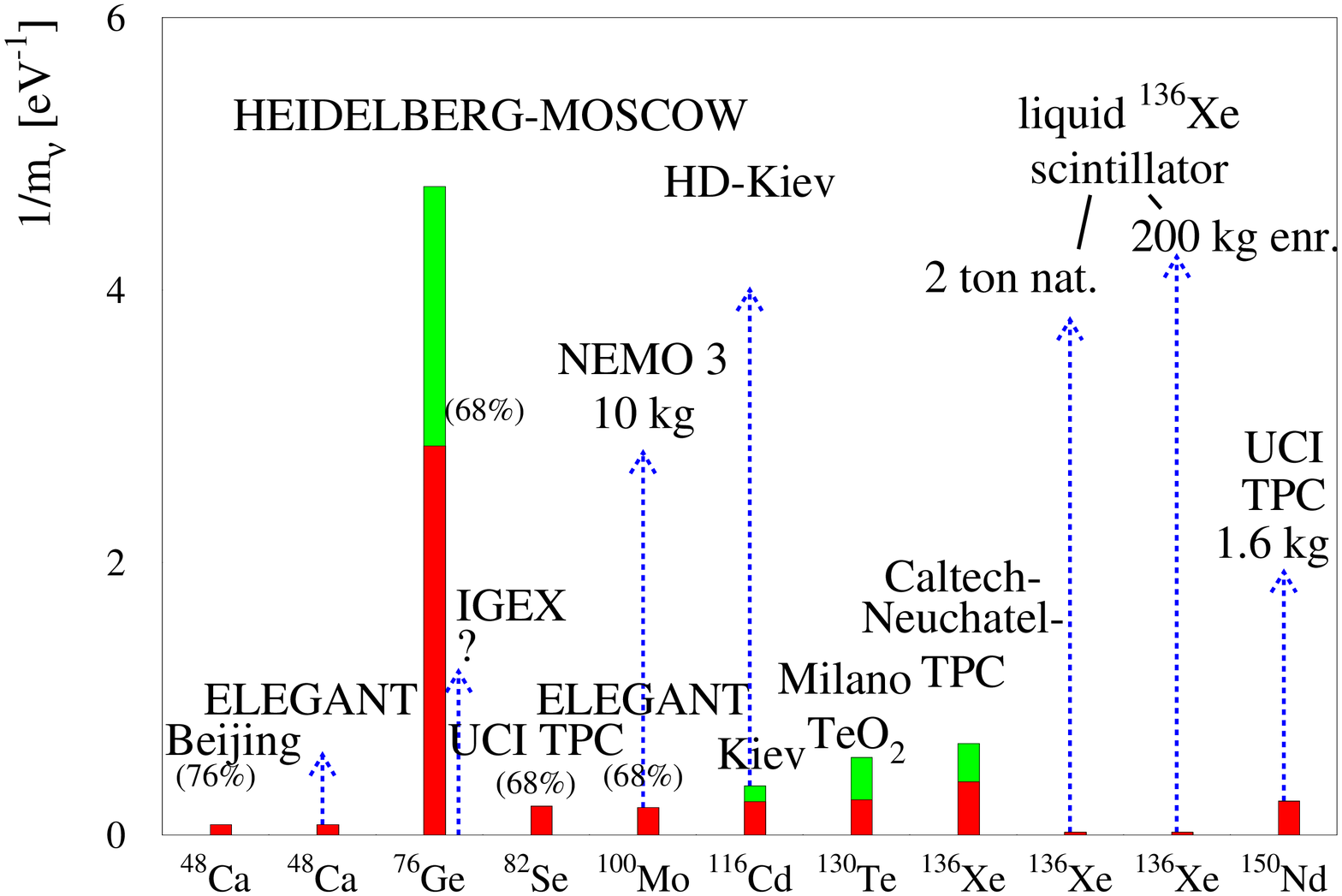,height=6cm,width=7cm}}}
\caption{Sensitivity of ${\beta \beta}_{0\nu}$ experiments. }
\label{betabetafut}
\end{figure}

If neutrinos are of Majorana type a new form of nuclear double beta
decay would take place in which no neutrinos are emitted in the final
state, i.e. the process by which an $(A,Z-2)$ nucleus decays to $(A,Z)
+ 2 \ e^-$. Such process  would arise from a virtual exchange of
Majorana neutrinos. Unlike ordinary double beta decay, the
neutrino-less process violates total lepton number L and its existence
would indicate the Majorana nature of neutrinos.  Because of the phase
space advantage, this process is a very sensitive probe of
the nature of neutrinos.  Present limits as well as future prospects
are illustrated in \fig{betabetafut}, taken from ref. \cite{Klapdor}.
Note that this bound depends to some extent on the relevant nuclear
matrix elements characterising this process \cite{haxtongranada} and
that the effective particle physics parameter $\VEV{m}$ involves both
neutrino masses and mixings. Thus, although rather stringent, this
limit may allow very large individual \neu masses, as there may be
strong cancellations between different neutrino types. These could
follow from suitable symmetries. For example, the decay vanishes if
the intermediate neutrinos are Dirac-type, as a result of the
corresponding lepton number symmetry \cite{QDN}.

Neutrino-less double beta decay has a great conceptual importance. It
has been first shown in ref. \cite{BOX} that in a gauge theory of the
weak interactions a non-vanishing ${\beta \beta}_{0\nu}$ decay rate
requires \neus to be Majorana particles, {\sl irrespective of which
mechanism} induces it. A recent application of this theorem to
supersymmetry was discussed here by Hirsch \cite{hirsch}.

\section{The Cosmological Density Limit }
\vskip .2cm

The oldest cosmological bound on neutrino masses follows from avoiding
the overabundance of relic neutrinos \cite{KT} 
\beq 
\label{RHO1}
\sum m_{\nu_i} \lsim 92 \: \Omega_{\nu} h^2 \: eV\:, 
\eeq 
where $\Omega_{\nu} h^2 \leq 1$ and the sum runs over all species of
isodoublet neutrinos with mass less than $O(1 \: MeV)$. Here
$\Omega_{\nu}=\rho_{\nu}/\rho_c$, where $\rho_{\nu}$ is the neutrino
contribution to the total density and $\rho_c$ is the critical
density.  The factor $h^2$ measures the uncertainty in the present
value of the Hubble parameter, $0.4 \leq h \leq 1$, and $\Omega_{\nu}
h^2$ is smaller than 1.  For the $\nu_{\mu}$ and $\nu_{\tau}$ this
bound is much more stringent than the laboratory limits \eq{1}.

Apart from the experimental interest, an MeV tau neutrino is
interesting from the point of view of structure formation \cite{ma1}.
Such \nt masses are theoretically viable as the constraint in
\eq{RHO1} holds only if \neus are stable on  cosmological
time scales. In models with spontaneous violation of total lepton
number \cite{CMP} neutrinos can decay into a lighter \neu plus a
majoron, for example \cite{fae},
\beq
\label{NUJ}
\nu_\tau \ra \nu_\mu + J \:\: .
\eeq
or have sizeable annihilations to these majorons,
\beq
\label{JJ}
\nu_\tau + \nu_\tau \ra J + J \:\: .
\eeq

The possible existence of fast decay and/or annihilation channels
could eliminate relic neutrinos and therefore allow them to have
higher masses, as long as the lifetime is short enough to allow for an
adequate red-shift of the heavy neutrino decay products. These 2-body
decays can be much faster than the visible modes, such as radiative
decays of the type $\nu' \ra \nu + \gamma$, which would also be highly
constrained by astrophysics and cosmology (for a detailed discussion
see ref. \cite{KT}).

A general method to determine the Majoron emission decay rates of
neutrinos was first given in ref. \cite{774}. The resulting decay
rates are rather model-dependent and will not be discussed here.
Explicit neutrino decay lifetime estimates are given in
ref. \cite{Romao92,fae,V}.  The conclusion is that there are many ways
to make neutrinos sufficiently short-lived and that all mass values
consistent with laboratory experiments are cosmologically acceptable.

\section{The Cosmological Nucleosynthesis Limit}
\vskip .2cm

\begin{figure}[t]
\centerline{\protect\hbox{\psfig{file=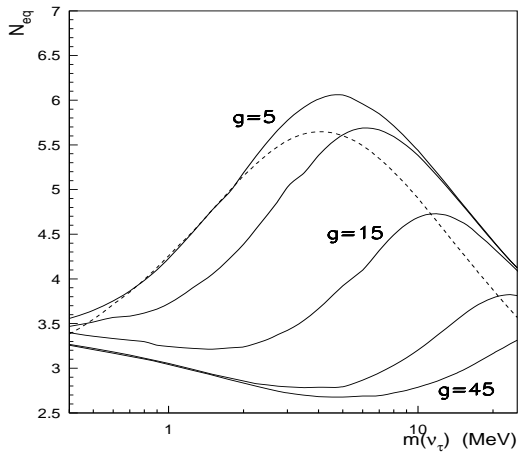,height=6cm,width=7cm}}}
\caption{A few MeV \nt annihilating to majorons can lower
the primordial helium abundance.}  
\label{neq} 
\end{figure}

\begin{figure}[t]
\centerline{\protect\hbox{\psfig{file=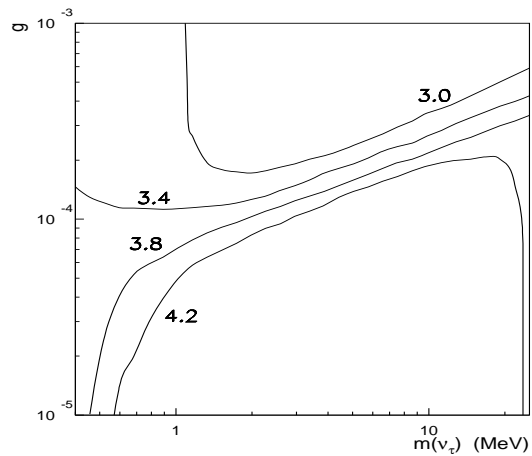,height=6cm,width=7cm}}}
\caption{The region above each curve is allowed for the
corresponding $N_{eq}^{max}$.}  
\label{neffmg}
\end{figure}

Stronger limits on neutrino lifetimes or annihilation cross sections
arise from cosmological nucleosynthesis. Recent data on the primordial
deuterium abundance \cite{dmeasure} have stimulated a lot of work on
the subject \cite{cris}.  If a massive \nt is stable on the
nucleosynthesis time scale, ($\nu_\tau$ lifetime longer than $\sim
100$ sec), it can lead to an excessive amount of primordial helium due
to its large contribution to the total energy density. This bound
can be expressed through an effective number of massless neutrino
species ($N_\nu$). If $N_\nu < 3.4-3.6$, one can rule out $\nu_\tau$
masses above 0.5 MeV \cite{KTCS91}.  If we take $N_\nu < 4$ the
\mnt limit loosens accordingly. However it has recently been argued
that non-equilibrium effects from the light neutrinos arising from the
annihilations of the heavy \nt's make the constraint a bit stronger in
the large \mnt region \cite{noneq}, ruling out all $\nu_\tau$ masses
oi this range.  One can show, however that in the presence of the \nt
annihilations in \eq{JJ} the nucleosynthesis \mnt bound is
substantially weakened or eliminated \cite{DPRV}.  Fig.~\ref{neq}
gives the effective number of massless neutrinos equivalent to the
contribution of a massive \nt for different values of the coupling $g$
between $\nu_\tau$'s and $J$'s, expressed in units of $10^{-5}$. For
comparison, the dashed line corresponds to the SM $g=0$ case. One sees
that for a fixed $N_\nu^{max}$, a wide range of tau neutrino masses is
allowed for large enough values of $g$. As long as $g$ exceeds a few
times $10^{-4}$ no \nt masses below the LEP limit can be ruled out.
One can express the above results in the $m_{\nu_\tau}-g$ plane, as
shown in figure \ref{neffmg}.  
One sees that the $\nu_\tau$ mass constraints from primordial
nucleosynthesis can be substantially relaxed, for values of
$g(m_{\nu_\tau})$ which are reasonable in many majoron models such as
those in ref. \cite{JoshipuraValle92,fae,DPRV,MASIpot3}.  Similar
depletion in massive \nt relic abundance also happens if the
\nt is unstable on the nucleosynthesis time scale as
will happen in many Majoron models \cite{unstable}.

\section{Supernova Limits  }
\vskip .2cm

There are a variety of limits on neutrino parameters that follow from
astrophysics \cite{Balantekin}, including the SN1987A observations, as
well as considerations from supernova (SN) theory, such as SN dynamics
\cite{Raffelt} and SN nucleosynthesis \cite{qian}. Here I briefly 
discuss some recent examples of how supernova physics constrains
neutrino parameters in various \neu conversion scenarios.

\begin{figure}[t]
\centerline{\protect\hbox{
\psfig{file=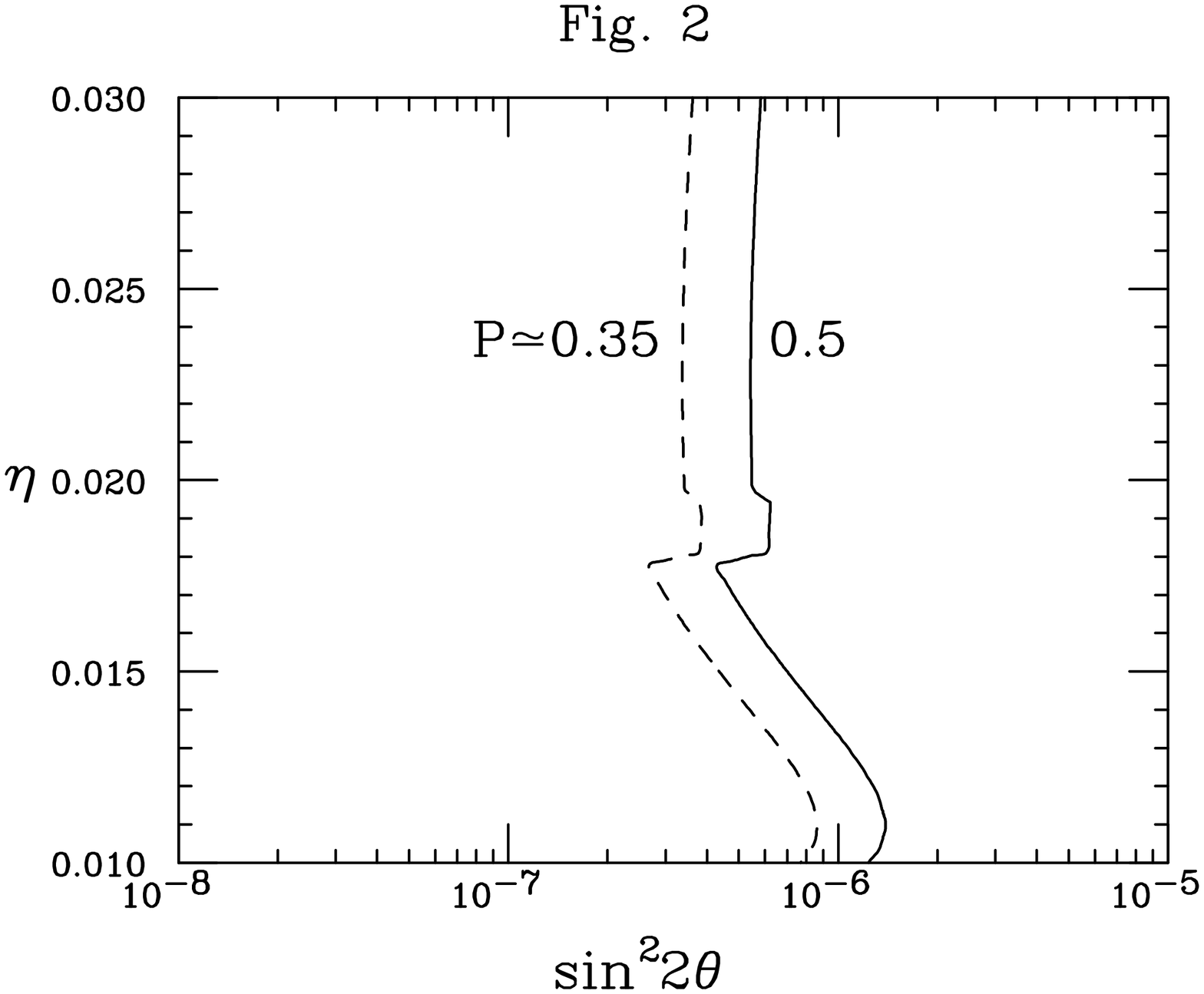,height=6cm,width=7cm,angle=90}
}}
\vglue -6.5cm
\hglue 2.7cm
\psfig{file=cover2.ps,height=1.2cm,width=3cm,angle=90}
\vglue 5cm
\caption{SN1987A bounds on massless neutrino mixing. }
\label{SN87}
\end{figure}

\begin{figure}[t]
\centerline{\protect\hbox{
\psfig{file=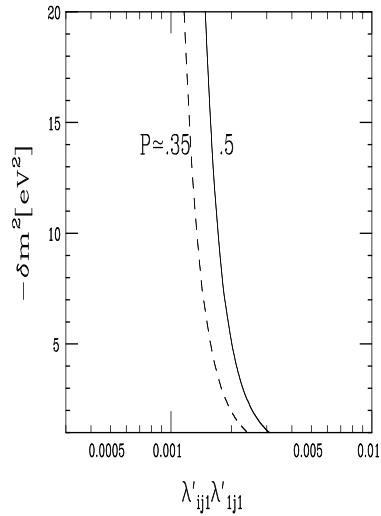,height=6cm,width=8cm,angle=90}
}}
\caption{SN1987A bound on R--parity-violating 
neutrino interactions. }
\label{fcncprob2}
\end{figure}

The observation of the energy spectrum of the SN1987A $\bar{\nu}_e$'s
may be used to provide very stringent constraints on the parameters
characterizing resonant neutrino conversions that could take place in
the dense supernova medium \cite{ssb}. This can be applied to various
\neu conversion scenarios in order to constrain the relevant
parameters. As an example, the regions to the right of the curves in
\fig{SN87}, taken from ref. \cite{massless}, are ruled out by SN1987A data,
while those to the left are allowed.  This is very surprising indeed,
as there is no way to constrain the mixing of massless \neus in the
laboratory, since no vacuum \neu oscillations are possible.
As another illustration consider the case of resonant \neu conversions
(massless or massive) induced by flavour changing neutral current
(FCNC) \neu interactions of broken R--parity models. The restrictions
that follow from the observed $\bar\nu_e$ energy spectra from SN1987A
are shown in \fig{fcncprob2}, taken from ref. \cite{rsusysn}, for the
case where \neu masses are in the hot dark matter range.

\begin{figure}[t]
\centerline{\protect\hbox{
\psfig{file=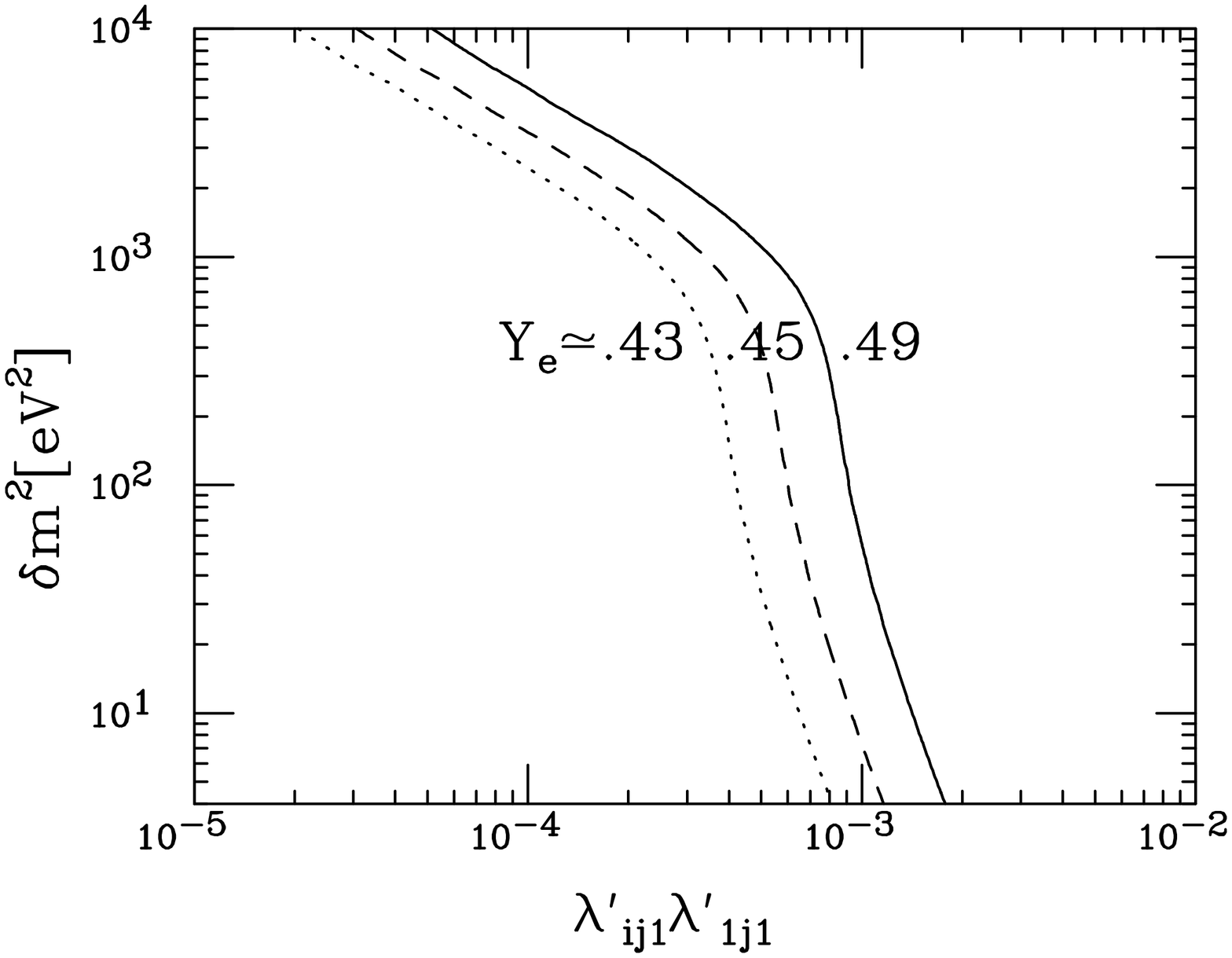,height=6cm,width=8cm,angle=90}
}}
\caption{Supernova r-process restriction on R--parity-violating 
neutrino interactions. }
\label{fcncye2}
\end{figure}

It has been noted that \neu conversions may play an important r\^ole
in SN $r$-process nucleosynthesis \cite{qian}.  This has been used in
order to obtain information on various \neu conversion scenarios.  In
\fig{fcncye2}, again from ref. \cite{rsusysn}, we display the
constraints on explicit $R$-parity-violating FCNCs for the opposite
sign of $\delta m^2$. Though this does not represent as solid a
constraint as that which follows from the SN1987A rate argument, it is
clear that it would be much more stringent than any constraint
obtained from the laboratory.  In particular, putting together the
results from \fig{fcncprob2} with those of \fig{fcncye2},
corresponding to two sign possibilities for $\delta m^2$, it is clear
that they disfavour a leptoquark interpretation of the recent HERA
anomaly.

\begin{figure}[t]
\centerline{\protect\hbox{
\psfig{file=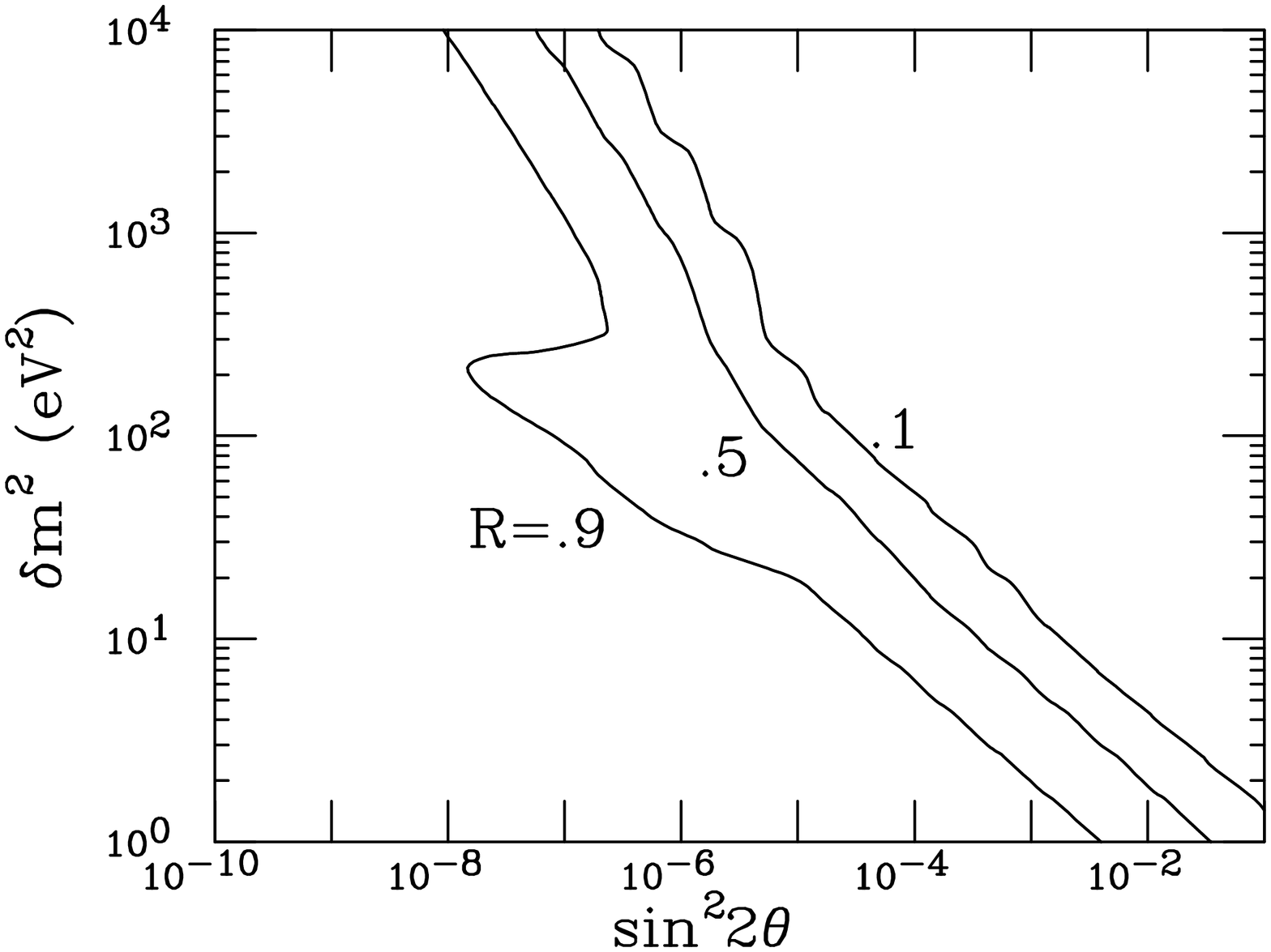,height=6cm,width=8cm,angle=90}}}
\caption{Supernovae and sterile neutrinos. }
\label{sterileSN}
\end{figure}

Another example of how astrophysics can constrain \neu properties
has to do with shock re-heating and the delayed explosion mechanism of
Bethe and Wilson. This has been recently considered within the scenario
of resonant $\nu_e \rightarrow\nu_s$ and $\bar{\nu}_e \to \bar{\nu}_s$
conversions in supernovae, where $\nu_s$ is a {\sl sterile} neutrino
\cite{nunus}, with mass in the hot dark matter mass range. In
\fig{sterileSN}, taken from ref. \cite{nunus}, the resulting
constraints on mixing and mass difference for the $\nu_e-\nu_s$ system
that follow from the supernova shock re-heating argument are
displayed. The implications of such a scenario for the detected
$\bar\nu_e$ signal from SN1987A and for the $r$-process
nucleosynthesis hypothesis have also been analysed in ref.
\cite{nunus}. An interesting result is that SN $r$-process
nucleosynthesis can be enhanced for some region of active-sterile
oscillation parameters.
As a final remark I mention a nice recent paper \cite{SNTM} showing
how resonant \neu conversions \cite{LAM} induced by Majorana \neu
transition moments \cite{BFD} can also be restricted by SN arguments.

In summary, neutrino masses now seem required in order to account for
the data on solar and atmospheric neutrinos, as well as the hot dark
matter component of the Universe. The scenarios needed in order to
reconcile these anomalies may have interesting implications
\cite{caprireview}. Detecting neutrino masses is one of the most
outstanding goals in particle physics, with far-reaching implications
also for the understanding of fundamental issues in astrophysics and
cosmology.  This work was supported by DGICYT grant PB95-1077 and by
EEC under the TMR contract ERBFMRX-CT96-0090.

\end{document}